# Photonic time-delayed reservoir computing based on lithium niobate microring resonators

YUAN WANG,[1] MING LI,[2] MINGYI GAO,[3] CHANG-LING ZOU,[2] CHUN-HUA DONG,[2] XIAONIU YANG,[1,4] QI XUAN,[1,4] AND HONGLIANG REN[1,4,*]

[1] *College of Information and Engineering, Zhejiang University of Technology, Hangzhou 310023, China*
[2] *CAS Key Laboratory of Quantum Information, University of Science and Technology of China, Hefei, Anhui 230026, China*
[3] *School of Electronic and Information Engineering School, Soochow University, Suzhou 215006, China*
[4] *Institute of Cyberspace Security, Zhejiang University of Technology, Hangzhou 310023, China*
\* *hlren@zjut.edu.cn*

**Abstract:** On-chip micro-ring resonators (MRRs) have been proposed for constructing delay reservoir computing (RC) systems, offering a highly scalable, high-density computational architecture that is easy to manufacture. However, most proposed RC schemes have utilized passive integrated optical components based on silicon-on-insulator (SOI), and RC systems based on lithium niobate on insulator (LNOI) have not yet been reported. The nonlinear optical effects exhibited by lithium niobate microphotonic devices introduce new possibilities for RC design. In this work, we design an RC scheme based on a series-coupled MRR array, leveraging the unique interplay between thermo-optic nonlinearity and photorefractive effects in lithium niobate. We first demonstrate the existence of three regions defined by wavelength detuning between the primary LNOI micro-ring resonator and the coupled micro-ring array, where one region achieves an optimal balance between nonlinearity and high memory capacity at extremely low input energy, leading to superior computational performance. We then discuss in detail the impact of each ring's nonlinearity and the system's symbol duration on performance. Finally, we design a wavelength-division multiplexing (WDM) based multi-task parallel computing scheme, showing that the computational performance for multiple tasks matches that of single-task computations.

## 1. Introduction

Recurrent Neural Networks (RNNs) are a specialized class of neural networks designed to handle sequential data by maintaining a memory of past inputs [1,2], thereby demonstrating superior performance in spatiotemporal tasks. However, RNNs are characterized by complexities such as slow training speeds and the challenges of vanishing or exploding gradients due to the intricacies of backpropagation through time (BPTT) [3]. Reservoir Computing (RC) is a more recent computational paradigm within RNNs that offers a reduction in training complexity compared to traditional RNNs and other neural network approaches [4,5]. The RC architecture consists of an input layer where data is assigned random and fixed weights before being transmitted to the reservoir. Within the reservoir layer, data is mapped to a higher-dimensional space through nonlinear nodes that are randomly and fixedly connected. The weights of the output layer are then typically trained using ridge regression or linear regression to address specific tasks. This unique characteristic of training only the output layer significantly reduces the training time of this neural network [5–7]. RC has shown promising applications in time series prediction, channel equalization, speech recognition, as well as various medical imaging and financial applications [8–10].

Photonic Neural Networks (PNNs) represent the forefront of AI computation, leveraging the unique properties of light, such as high bandwidth, low latency, and the potential for low power consumption [11]. Numerous studies have demonstrated the implementation of photonic RC, where nonlinear nodes are typically realized either through the nonlinear behavior of photonic devices [12,13] or the dynamics of nonlinear optical phenomena [14,15]. The first approach, known as spatial reservoir computing, is quite similar to RNNs, where the nodes are spatially distributed and can establish physical connections. Typical examples of spatial reservoirs include networks constructed from semiconductor optical amplifiers and networks where each pixel of a spatial light modulator serves as a node [18]. However, this method employs photonic device modules as nonlinear nodes [16], which presents challenges in terms of scalability and significantly increases the footprint of the photonic circuitry. An alternative approach, known as Time-Delay Reservoir Computing (TDRC), involves temporal multiplexing of nodes, utilizing a single physical nonlinear node that typically receives feedback through a physical loop to enhance connectivity and memory among virtual nodes [17]. Most proposed RC schemes are designed using passive integrated optical components based on SOI (Silicon-On-Insulator) technology, including photonic waveguides [18–20], photonic crystal microcavities [21,22], and silicon-based microring resonators (MRRs) [23,24]. The RC schemes based on photonic waveguides are widely favored due to their lower design complexity and higher tolerance to errors, though the large footprint of waveguides poses challenges for integration. In contrast, photonic crystal microcavity-based RC schemes offer significant advantages in terms of footprint and stronger light confinement, holding great potential, but they demand extremely high fabrication precision. Silicon-based MRRs, however, perform well in terms of light manipulation, size, and fabrication error tolerance, though the nonlinear capability of silicon might be somewhat limited. Given the advantages and disadvantages of these schemes, exploring materials with greater nonlinear potential could be a promising research direction. However, RC schemes based on Lithium Niobate on Insulator (LNOI) have not yet been reported. In these RC systems, nonlinearity is typically introduced through square-law photodetectors or the nonlinear response of resonators [31,32]. Memory is determined by delay lines connecting the nodes [18] or the photon lifetime within microcavities [32]. In Ref. [32], to ensure that the RC system has sufficient memory capacity (MC) to perform specific computational tasks, a single silicon-based MRR combined with an external optical feedback waveguide was employed to construct the RC system. The inclusion of the optical feedback waveguide significantly increased the system's linear MC. However, the optimal length of the feedback waveguide is approximately 20 cm, far exceeding the diameter of the microring, which introduces significant challenges in device fabrication, transmission loss, and temperature control, hindering the integration and scalability of photonic RC systems. To address this issue, in our previous work, we proposed a method to construct a time-delay RC system using a silicon-based main cavity coupled with a series of resonator arrays [33]. The series-coupled resonator arrays can replace the previously overly long optical feedback waveguide.

As one of the most widely used synthetic crystals, Lithium Niobate (LN) plays a crucial role in modern telecommunications due to its exceptional electro-optic and optical $\chi^2$ nonlinearity, as well as its broad optical transparency window [2,25,26]. Of particular interest is the complex nonlinear oscillatory behavior of intracavity optical waves that arises when a competing mechanism, such as the photothermal nonlinear effect, occurs within the resonator [27]. LN microphotonic devices enable the exploration of low-energy photonic computation [28–30]. In this paper, we build upon this foundation to investigate and evaluate the performance of a RC scheme based on LNOI microring design. We further extend the study to examine scenarios involving the detuning of the resonant wavelength between the additional series-coupled cavity array and the main cavity, the nonlinearity of each ring in the additional cavity array, the time duration of bit periods, and multi-task parallel computing. In the proposed reservoir design, a single main cavity serves as the physical nonlinear node, while the series-coupled resonators further contribute to the formation of delayed feedback loops. This configuration significantly enhances the system's memory capacity. We conducted numerical analysis and evaluation of the proposed RC system's performance on the discrete-time tenth-order nonlinear autoregressive moving average (NARMA-10) task. The results indicate that the proposed system significantly improves computational performance when the memory required by the task exceeds what a single microring resonator (MRR) can provide. We numerically analyzed and evaluated the performance of the proposed RC system under these four conditions.

## 2. Principle

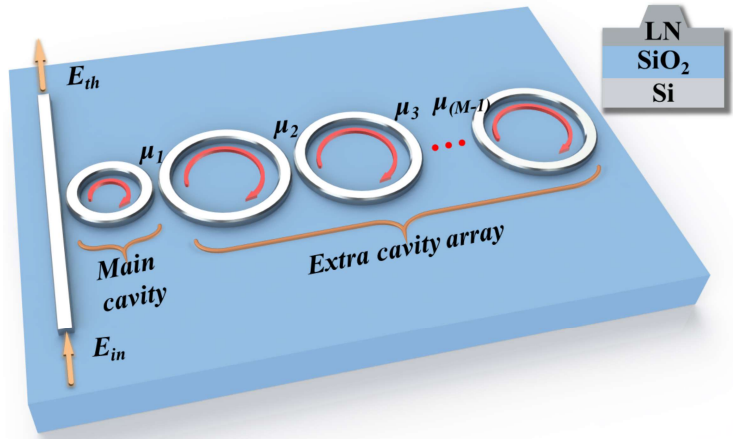

Fig. 1. Schematic diagram of SCMRRs designed for constructing time-delayed reservoir computing system, which consists of a LNOI -based main cavity coupled in series with an extra microring resonator array.

Figure 1 illustrates the configuration of a series-coupled microring resonator (SCMRR) based on LNOI used for constructing a TDRC system. This configuration, designed as an all-pass filter structure, consists of a waveguide and a series of MRRs that are side-coupled to the waveguide. In this series-coupled arrangement, all MRRs are fabricated from LNOI. For clarity, the MRR directly coupled to the waveguide is referred to as the main cavity, while the others are designated as the auxiliary cavity array. In this section, we provide a comprehensive model of the LNOI-based SCMRR, establishing it as a fundamental component of an optical reservoir computing architecture. Our analysis is grounded in a theoretical framework that utilizes coupled mode theory (CMT) [34,35,36,37].

The MRR structure, as one of the commonly used devices for wavelength-division multiplexing (WDM), also serves as an inspiration for the proposed multi-task parallel computing scheme [42]. The $l^{th}$ resonance frequency of the MRR depends on its radius $R$ and the effective refractive index $n_{eff}$ of the propagating mode, as expressed in Eq. (1). Here, $c$ represents the speed of light in a vacuum.

$$\omega_{1,i} = \frac{c}{n_{eff}R} \cdot l, \quad l = 1, 2, 3, \ldots \tag{1}$$

The wavelength range between two resonances is commonly referred to as the Free Spectral Range (FSR), which can be expressed in terms of the resonance wavelength and the group index $n_g$:

$$FSR \approx \frac{\lambda^2}{2\pi R n_g} \quad (2)$$

In the designed SCMRRs system, the radius of the primary cavity is half that of the auxiliary cavity, resulting in the primary cavity's FSR being twice that of the auxiliary cavity. Consequently, when the resonance frequency of the primary cavity is used as the corresponding task channel, a matching frequency channel can be found at the resonance frequency of the auxiliary cavity. At this point, each frequency channel within the SCMRRs system can modulate the corresponding task input signal, enabling multi-task parallel computing based on wavelength-division multiplexing. For $n$ task channels, equations (3) to (7) provide the nonlinear dynamical equations governing the time evolution of three state variables [27, 38–40].

$$\frac{dU_{1,i}(t)}{dt} = [i(\omega_{1,i}(t) - \omega_{p,i}) - \Gamma_t/2]U_{1,i}(t) + i\mu E_{in,i}(t) + i\mu_1 U_{2,i}(t) \quad (3)$$

$$\frac{dU_{m,i}(t)}{dt} = [i(\omega_{m,i}(t) - \omega_{p,i}) - \Gamma_{tm}/2]U_{m,i}(t) + i\mu_{m-1}U_{m-1,i}(t) \\ + i\mu_{m+1}U_{m+1,i}(t), (2 \leq m \leq M-1) \quad (4)$$

$$\frac{dU_{M,i}(t)}{dt} = [i(\omega_{M,i}(t) - \omega_{p,i}) - \Gamma_{tM}/2]U_{M,i}(t) + i\mu_{M-1}U_{M-1,i}(t) \quad (5)$$

$$\frac{d\overline{E}_{sp,m,i}(t)}{dt} = -\Gamma_E \overline{E}_{sp,m,i} + \eta_E \sum_{i=1}^{n}|U_{m,i}(t)|^2, 1 \leq m \leq M \quad (6)$$

$$\frac{d\Delta\overline{T}_{m,i}(t)}{dt} = -\Gamma_T \Delta\overline{T}_{m,i} + \eta_T \sum_{i=1}^{n}|U_{m,i}(t)|^2, 1 \leq m \leq M \quad (7)$$

Equation (3) describes the time evolution of the optical power amplitude $U_{1,i}$ in the primary cavity within the $i$th optical frequency channel, considering its coupling with the waveguide input field amplitude $E_{in,i}(t)$ and the optical power amplitude $U_{2,i}$ of the secondary cavity adjacent to the primary cavity. Here, $\omega_{p,i}(t)$ is the angular frequency of the incident light in the $i$th optical frequency channel, $\omega_{1,i}(t)$ is the corresponding resonance angular frequency of the primary cavity, $\Gamma_t$ represents the loss rate, and $\mu$ and $\mu$ are the energy coupling coefficients between the primary cavity and the straight waveguide or the nearest secondary cavity adjacent to the primary cavity, respectively. Equations (4) and (5) describe the time dynamics of the complex optical power amplitude in the secondary cavities. In this series, the secondary cavities are indexed from 2 to $M$, where the nearest secondary cavity to the primary cavity is labeled as 2, and the farthest secondary cavity is labeled as $M$. Equation (6) describes the dynamical evolution of the space charge field $\overline{E}_{sp}$ in each cavity, which originates from the photovoltaic current generated by the asymmetric excitation of carriers due to light absorption, with a rate $\eta_E$ and a recombination process governed by the decay constant $\Gamma_E$. Equation (7) describes the variation in the modal average temperature $\Delta T$ within each cavity, where $\Gamma_T$ is the thermal relaxation rate, and $\eta_T$ is the photothermal heating coefficient related to light absorption and heat conversion.

$\mu$ and $\mu_m$ represent the coupling between the waveguide and the primary cavity, and the coupling between the micro-ring cavities, respectively [35].

$$\mu^2 = \frac{\kappa^2 c}{2\pi n_{g1} R_1} \tag{8}$$

$$\mu_m^2 = \frac{\kappa_m^2 c^2}{(2\pi n_{gm} R_m)(2\pi n_{g(m+1)} R_{m+1})} \tag{9}$$

Here, $\kappa^2$ denotes the power coupling between the primary cavity and the waveguide, while $\kappa_m^2$ ($m = 1,2,3, \cdots, M$-1) represents the power coupling coefficient between the $m^{th}$ micro-ring cavity and the $(m+1)^{th}$ cavity. $c$ represents the speed of light in a vacuum. $n_{gm}$ and $R_m$ denote the group refractive index and the radius of the $m^{th}$ micro-ring cavity, respectively. For simplicity, this paper assumes that all secondary cavities are identical. Therefore, the coupling coefficients between any two adjacent secondary cavities are identical ($\mu_2 = \mu_3 = \ldots = \mu_{M-1}$), and all secondary cavities share the same resonance frequency ($\omega_2 = \omega_3 = \ldots = \omega_M$), group index ($n_{g2} = n_{g3} = \ldots = n_{gM}$), radius ($R_2 = R_3 = \ldots = R_M$), and quality factor ($Q_2 = Q_3 = \ldots = Q_M$). The output electric field signal $E_{th,i}$ of the $i^{th}$ optical frequency channel can be expressed as:

$$E_{th,i}(t) = t_r E_{in,i}(t) + \sqrt{\Gamma_e} U_{1,i}(t) \tag{10}$$

In the equation, $t_r$ represents the field transmission from the input port to the through port ($t_r^2 + \kappa^2 = 1$).

The thermo-optic nonlinearity and the photorefractive effect are the two primary mechanisms responsible for the nonlinear dynamics observed in LN micro resonators [38,39]. On one hand, the response of thermo-optic nonlinearity is directly related to the dynamics of the device temperature under photothermal heating. On the other hand, the photorefractive effect in LN is essentially an electro-optic effect caused by the space charge electric field generated by the photovoltaic drift current. Consequently, the resonant frequency $\omega_1$ of the main cavity depends on both the device temperature and the space charge electric field [39,40].

$$\omega_1(\Delta \overline{T}_1, \overline{E}_{sp,1}) = \omega_{c1} + g_T \Delta \overline{T}_1 + g_E \overline{E}_{sp,1} \tag{11}$$

Here, $\omega_{c1}$ denotes the cold-cavity resonant frequency of the main cavity in the absence of thermo-optic nonlinearity and the photorefractive effect. $\Delta \overline{T}_1$ and $\overline{E}_{sp,1}$ represent the average device temperature variation and the average space charge electric field over the optical mode field profile, respectively [39,40]. $g_T \equiv \frac{d\omega_1}{dT}$ is the photothermal coupling coefficient, and $g_E \equiv \frac{d\omega_1}{dE}$ is the electro-optic coupling coefficient. The second term on the right side of Equation (11) represents the resonance frequency shift caused by the thermo-optic effect, where the photothermal coupling coefficient is denoted as

$g_T = -\omega_{c1}(\alpha_n + \alpha_l)$. Here, $\alpha_n$ represents the thermo-refractive effect, and $\alpha_l$ accounts for the thermo-expansive effect.

For simplicity, we assume that the thermo-refractive and thermo-expansive effects result in the same resonance frequency shift for a given temperature change. In an x-cut LN micro resonator, the quasi-TE polarized cavity modes, where the optical field propagates within the resonator (with in-plane dominant polarization), are likely to evolve into ordinary and extraordinary polarizations. Therefore, $\alpha_n$ can be well approximated as $\alpha_n \approx \frac{1}{2}(\frac{1}{n_e}\frac{dn_e}{dT} + \frac{1}{n_o}\frac{dn_o}{dT})$, where $n_o$ and $n_e$ represent the refractive indices of the ordinary and extraordinary polarizations, respectively [27]. Similarly, $\alpha_l$ is approximated as $\alpha_l \approx \frac{1}{2}(\alpha_l^{(Z)} + \alpha_l^{(X,Y)})$. The third term on the right side of Eq. (11) represents the electro-optic effect induced by the space charge electric field. In LN, the space charge electric field is primarily generated by the photovoltaic drift current along the crystal axis [41]. The results indicate that for quasi-TE polarized cavity modes in x-cut LN microresonators, a detailed analysis shows that the electro-optic coupling coefficient can be expressed as $g_E \approx \frac{n_o^2}{4}\omega_{c1}(r_{13} + r_{33})$, where $r_{13}$ and $r_{33}$ are the electro-optic coefficients of LN [27].

When the primary cavity operates in a linear state, we can establish the conditions $g_T = 0$ and $g_E = 0$. Therefore, in the absence of nonlinearity, the characteristic time scale of the primary cavity is determined by its photon lifetime $\tau_{ph} = \Gamma_t^{-1}$. Since the system's nonlinear state is induced by the thermo-optic nonlinear effect and the photorefractive effect, the two key time scale parameters $\tau_T = \Gamma_T^{-1}$ and $\tau_E = \Gamma_E^{-1}$ are related to the system's nonlinear dynamical evolution. Parameter $\tau_T$ is two orders of magnitude smaller than parameter $\tau_E$. The nonlinear effects will influence the dynamical evolution only when the time scale of the input signal matches the specific time scale parameter. In this study, the bit duration of the input signal is adjusted to align with the time scale of $\tau_T$, thus we particularly emphasize the nonlinear effects caused by the presence of thermo-optic phenomena in the primary cavity.

**Table 1. Parameters values used in simulation**

| Parameter | Value | Parameter | Value |
| --- | --- | --- | --- |
| $R_1$ | 45 μm | $\alpha_l(X,Y)$ | $0.748 \times 10^{-5}$ |
| $R_2 \sim R_M$ | 90 μm | $r_{13}$ | 8.6 pmV$^{-1}$ |
| $\lambda_0$ | 1.5111 μm | $r_{33}$ | 30.8 pmV$^{-1}$ |
| $n_0$ | 2.21 | $\Gamma_T$ | 10Mhz |
| $n_e$ | 2.14 | $\Gamma_E$ | 100Khz |
| $\alpha_l(z)$ | $1.538 \times 10^{-5}$ | $\eta_T$ | $3.75 \times 10^{17}$ |

For the study of the system's dynamics, we define the initial wavelength detuning between the laser wavelength and the MRR resonance as $\Delta\lambda_s = \lambda_p - \lambda_0$, and the resonance shift caused by nonlinear effects as $\Delta\lambda_0(t) = \lambda_0(t) - \lambda_0$, where $\lambda_0(t) = 2\pi c/\omega_0(t)$ and $\lambda_0 = 2\pi c/\omega_0$. The set of coupled differential equations, Eq. (1)–(10), is numerically integrated using the Runge-Kutta method with a time step of 20 ps, which is sufficient to capture the effects of the smallest time scale ($\tau_{ph}\approx$1845 ps). The model considers only unidirectional propagation, and the parameter values are reported in Table 1. For the remainder of this paper, to simplify the nomenclature, configurations with different numbers of serially coupled auxiliary cavities are denoted by a simplified version. For example, if the SCMRRs system has two auxiliary cavities, it is referred to as SCMRRs-2.

## 3. Constructing time-delayed RC with SCMRRs

A TDRC system was constructed using the SCMRR structure. Figure 2 illustrates a schematic of the proposed four-frequency-channel TDRC system, which includes an input layer, a reservoir layer, and an output layer [33]. The number of enabled frequency channels is determined by the number of parallel tasks. When only one task is performed, only the channel operating at the frequency $\omega_{1,0}$ is used, whereas all frequency channels are utilized when performing four tasks. The resonance frequencies of the primary cavity correspond one-to-one with those of the auxiliary cavity array, and the wavelength detuning of the laser operating in each frequency channel is kept consistent to facilitate subsequent optimization. Taking a single channel as an example, in the input layer, the time-continuous input signal is first encoded into a bit sequence, where $x_i$ represents the amplitude of the $i^{th}$ bit, and $\tau$ represents the bit period. A bit mask is then applied by multiplying the bit stream by a set of random values $M(t)$, resulting in $S(t)$. The mask set $M(t)$ has a size of $N_v$, with elements following a uniform distribution. A bias $\beta$=8.0 is then added to $S(t)$, a value optimized for RC performance (with all other parameters fixed). The resulting signal is modulated onto the intensity of an optical carrier, with the maximum laser input power denoted as $P_M$. The modulated optical signal then enters the SCMRR system through the input port and propagates through all MRRs via their coupling. A photodetector (PD) connected to the through port converts the received optical signal into an electrical signal. Within a single $\tau$, the electrical signal is synchronously sampled at a specified masking interval $\theta$. By configuring the sampling interval to satisfy $\theta = \tau/N_v$, $N_v$ equidistant sampling points are uniformly distributed over time. This ensures that the number of sampling points precisely matches the size of the mask set. The $N_v$ equidistant points act as $N_v$ virtual nodes, which function similarly to nodes in a conventional reservoir [43]. In the reservoir layer, both the primary cavity and the auxiliary cavities exhibit nonlinear characteristics, causing the input signal to undergo nonlinear transformations. A series-coupled auxiliary cavity array is designed to enhance the MC. As a result, the initial input signal bits undergo nonlinear transformations, transitioning from the physical system to a high-dimensional space characterized by $N_v$ virtual nodes. Finally, in the output layer, a linear combination of the responses from the relevant virtual nodes yields the predicted value $O_i$ corresponding to the input $x_i$, as shown below:

$$O_i = \sum_{j=1}^{N} W_j \mathrm{X}_{j,i} \qquad (12)$$

In the equation, $\mathrm{X}_{j,i}$ represents the elements of an nnn-dimensional vector, indicating the response of the virtual nodes during the iiith period, and $W_j$ denotes the corresponding readout weights. The weights in the readout layer are optimized using ridge regression to minimize the normalized mean squared error (NMSE) between the predicted value $O_i$ and its desired corresponding value $y_i$ [32], which is expressed as:

$$NMSE = \frac{\langle \|O_i - y_i\|^2 \rangle}{\langle \|y_i - \langle y_i \rangle\|^2 \rangle} \quad (13)$$

The intrinsic and loaded quality factors of the primary cavity are $8.3 \times 10^6$ and $2.3 \times 10^6$, respectively. The competition between thermo-optic and photorefractive nonlinearities in LN MRRs may lead to self-pulsing phenomena. The masking interval $\theta=200$ps is significantly shorter than the photon lifetime ($\tau_{ph} \approx 1845$ps), thermal lifetime ($\tau_T \approx 100$ns), and space charge field lifetime ($\tau_E \approx 10$μs), with the system's nonlinearity linked to the latter two timescales. The short masking interval, being less than the photon lifetime, ensures that the primary cavity's internal field persists between consecutive signals, preserving prior state information and facilitating coupling between adjacent virtual nodes—a phenomenon known as short-term memory [32]. Introducing a series-coupled auxiliary cavity array enhances MC by extending signal retention, thus preserving earlier state information. The nonlinear behavior of the auxiliary cavities, which improves system performance, is further examined in Section 4. With a bit duration of $\tau = 10$ns, the thermo-optic nonlinearity becomes more prominent, enabling faster computation due to its quicker time response. This results in $N_v=50$ virtual nodes, which will be used consistently in subsequent simulations. The modulated optical signal propagates through the primary cavity, inducing thermo-optic effects and causing a resonance shift $\Delta\lambda_T$. Simultaneously, the photorefractive effect causes a resonance shift $\Delta\lambda_E$. However, due to the signal's rapid timescale relative to ($\tau_T$, $\tau_E$), these shifts mainly cause wavelength detuning rather than nonlinear signal transformation. High optical power in the high-Q primary cavity can lead to significant self-pulsing dynamics, dominated by thermal effects.

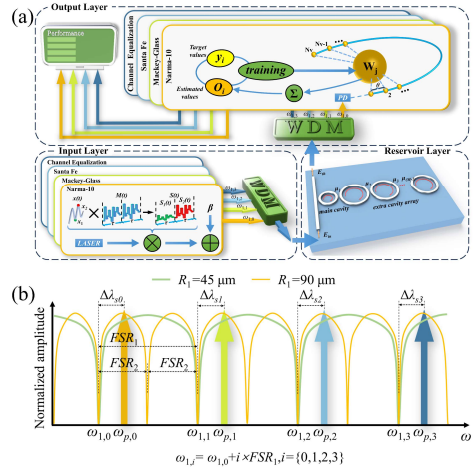

Fig. 2 (a) Schematic diagram of the multi-task time-delay RC system with SCMRRs configuration. The input information $x_t$ is first masked by a random sequence $m(t)$, followed by the addition of a bias $\beta$. The resulting electrical signal is then modulated onto the intensity of an optical carrier, which is excited by a laser. At the through port, the optical signal is converted into a corresponding electrical signal by a photodetector (PD). Virtual nodes in the reservoir are created through time-division multiplexing, and the predicted value $O_i$ is obtained by a linear weighted sum of the responses from the relevant virtual nodes. The weights are optimized using a linear classifier with supervised learning on the expected values of the dataset. (b) Typical linear transmission responses of MRRs with radii of 45 μm and 90 μm, along with the frequency channel allocation used in this work. PD: Photodetector.

## 4. Benchmark

### 4.1 NARMA 10

The NARMA-10 chaotic time series prediction task requires the system's nonlinear response and memory capacity, relying on information from the previous 10 time steps. The target equation for the NARMA-10 task is as follows:

$$y_{i+1} = 0.3y_i + 0.05y_i \sum_{k=0}^{9} y_{i-k} + 1.5x_{i-9}x_i + 0.1 \quad (14)$$

Here, $x_i$ is the random input at time $i$, uniformly distributed within [0, 0.5], and $y_i$ is the corresponding output.

### 4.2 Mackey-Glass

The Mackey-Glass time series is a standard benchmark for chaotic time series prediction tasks [44]. This series is defined by the following differential equation:

$$\frac{dy(t)}{dt} = \frac{0.2y(t-\tau)}{1+y(t-\tau)^{10}} - 0.1y(t) \quad (15)$$

In this task, $y(t)$ is the output at time $t$ with a delay $\tau$, where $\tau=17$.

### 4.3 Santa Fe

The Santa Fe laser time series task involves one-step prediction of chaotic far-infrared laser intensity. This task requires sufficient memory capacity (MC) from a single MRR and moderate system nonlinearity. Compared to NARMA-10 and Mackey-Glass tasks, the Santa Fe dataset includes experimental noise, adding extra challenges for accurate prediction.

### 4.4 Channel Equalization

In the wireless channel equalization task, the signal propagates through a noise-affected wireless channel, which is modeled as a linear system with second- and third-order nonlinear distortions due to multipath propagation. The original signal $d(n)$ is an independent, identically distributed sequence {-3, -1, 1, 3}. The output of the linear wireless channel model, $q(n)$, is given by:

$$\begin{aligned} q(n) = &\ 0.08d(n+2) - 0.12d(n+1) + d(n) + 0.18d(n-1) - 0.1d(n-2) \\ &+ 0.091d(n-3) - 0.05d(n-4) + 0.04d(n-5) + 0.03d(n-6) \\ &+ 0.01d(n-7) \end{aligned} \quad (16)$$

The output sequence of the system, $u(n)$, is affected by noise $v(n)$ and higher-order nonlinear distortions, defined as:

$$u(n) = q(n) + 0.036q^2(n) - 0.011q^3(n) + v(n) \quad (17)$$

During training, the squared error between the reconstructed and original signals is minimized. In post-processing, the RC output is rounded to the nearest symbol in {-3, -1, 1, 3}, generating the reconstructed signal $y(n)$. The symbol error rate (SER) between the original and reconstructed signals is calculated at a signal-to-noise ratio (SNR) of 24 dB.

## 5. Results

Under the conditions described above, the SCMRRs system was analyzed in detail, focusing on the detuning of resonance wavelengths between the auxiliary cavity array and the primary cavity, the nonlinearity of each ring in the auxiliary array, symbol duration, and multi-task parallel computing.

In photonic RC, the importance of nonlinear signal transformation is undeniable, but it can also disrupt MC [45]. To optimize performance, RC systems must balance nonlinear transformations with MC. However, measuring the appropriate level of nonlinearity and required MC for specific tasks is challenging. The former can be indirectly evaluated by the standard deviation of the resonance wavelength shift $\sigma(\Delta\lambda_0(t))$, with higher values indicating increased nonlinearity. The latter is assessed by examining linear MC, where the reservoir is trained to accurately reconstruct an input stream from 0 to 0.5 after $k$ time steps. It is calculated as:

$$MC = \sum_{k=1}^{l_{max}} MC_k \tag{18}$$

$$MC_k = \frac{\text{cov}^2(x_{i-k}, y_k)}{\text{var}(x_i)\text{var}(y_k)} = 1 - NMSE \tag{19}$$

Here, $MC_k \in [0,1]$ represents the $MC$ at displacement $k$, with $MC_k = 1$ indicating perfect retention of the bit stream after $k$ steps, and $MC_k = 0$ indicating complete memory loss. $l_{max}$ denotes the maximum length of the memory sequence, while var(·) and cov²(·) represent variance and covariance, respectively.

Then, we optimized four key parameters to ensure the best performance of the SCMRRs system in each scenario: maximum input laser power $P_M$, initial wavelength detuning $\Delta\lambda$s, the $Q_1/Q_0$ ratio, and the total number of MRRs $M$. The primary cavity's nonlinear dynamics are influenced by $P_M$ and $\Delta\lambda$s, while the system's memory capacity (MC) is affected by the number and quality factor of the auxiliary cavities. The primary cavity has a radius of 45 μm, while the auxiliary cavities have a radius of 90 μm and equal or greater quality factors than the primary cavity. $P_M$ ranges from 1 μW to 200 μW, and $\Delta\lambda$s is adjusted from -3 pm to 3 pm, with a step size of 0.1 pm, covering the primary cavity's resonance FWHM of 0.66 pm. The $Q_1/Q_0$ ratio is varied from 1 to 10 by changing $Q_1$ while keeping $Q_0$ constant. $M$ ranges from 1 (only the primary cavity) to 10 (SCMRRs-9 with 9 auxiliary cavities). During evaluation, the first 1000 data bits are used to preheat the system, 2000 bits for training, and the next 1000 for testing, with distinct data for training and testing. All simulations use the same random mask set, uniformly distributed between 0 and 1. A linear classifier with ridge regression coefficient $10^{-9}$ is used for the RC output layer. Unless stated otherwise, the study uses these parameter values.

### 5.1 Resonance wavelength offset between the auxiliary cavity array and the primary cavity

According to Ref.[33], optimal performance in the NARMA-10 task requires high MC without significant nonlinear transformation. For our RC model, extensive simulations show the best performance at $P_M$ = 0.06mW and $\Delta\lambda_s$ =1.2 pm, with laser power lower than in Ref.[33] due to material properties. At $P_M$ = 0.06mW and $\Delta\lambda_s$ =1.2 pm, we further optimize resonance wavelength offset, quality factor ratio, and MRR count, ensuring all MRRs operate nonlinearly.

Since initial wavelength detuning greatly impacts system performance, we thoroughly investigate the resonance wavelength offset, defined as offset=$\lambda_{main}$-$\lambda_{additional}$.

In Fig. 3(a), the parameter space for offset is divided into three regions based on NMSE performance: A, B, and C. Region A, near the MRR (cold) resonance, shows poor NMSE results (mostly >1) with no significant memory improvement. Region B, between A and C, achieves the best RC performance with NMSE of 0.0634, balancing nonlinearity and memory capacity. Region C, at higher detuning, results in the loss of memory enhancement from the auxiliary cavity array, showing similar performance to SCMRRs-0 with NMSE of 0.4456. Generally, higher MC correlates with lower NMSE. Figures 3(c) and 3(d) show normalized training output for offsets of 8 pm and 0 pm, respectively. Self-pulsing in the lithium niobate micro-rings at 0 pm offset indicates strong nonlinearity, leading to poor prediction. Figures 3(e) and 3(f) illustrate NMSE and MC variations with $Q_1/Q_0$ and MRR count $M$. MC and NMSE exhibit local fluctuations rather than continuous changes, similar to results from silicon micro-ring systems. This discontinuity arises from nonlinear instabilities when varying parameters, causing local oscillations in computational performance.

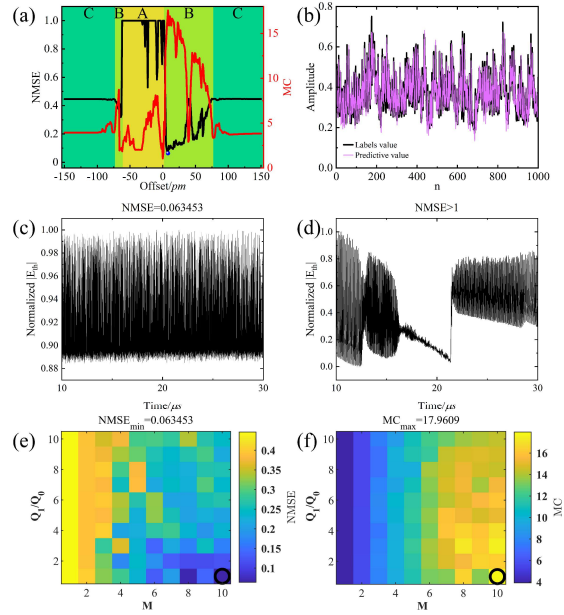

Fig. 3 (a) Under the conditions $M$=10 and $Q_1/Q_0$=1, the relationship between the resonance wavelength shift of the extra ring array and system performance. (b) Target (black) and prediction (purple) of NARMA-10 at the blue point. Normalized transmission waveforms of the through port under identical conditions: (c) offset = 8 pm and (d) offset = 0 pm. For a symbol duration of 10 ns, NMSE(e) and MC(f) versus the quality factor's ratio Q1/Q0 and the total number M of MRRs. And the black circles represent the conditions under which the system performs optimally.

## 5.2 the nonlinearity of each ring in the auxiliary array

In previous silicon-based SCMRR systems, only the primary cavity exhibited nonlinearity, simplifying simulations but potentially limiting performance. In this study, all micro-rings in the lithium niobate-based SCMRR system are set to exhibit nonlinear behavior. To explore the impact of nonlinear auxiliary cavities on system performance, we gradually remove the nonlinear constraints on each cavity in the SCMRRs-9 system. Nonlinear behavior arises from thermo-optic and photorefractive effects; when set to linear, both coupling coefficients $g_T$ and $g_E$ are zero, removing these effects.

In the SCMRRs-9 system with $Q_1/Q_0=1$, $P_M = 0.06$mW, $\Delta\lambda_s =1.2$ pm, and offset = 8 pm, nonlinear behavior in micro-rings significantly enhances performance (Figure 4(a)). When all rings are linear, the NMSE for the NARMA-10 task is 0.3262, better than SCMRRs-0 (NMSE = 0.4456) due to the auxiliary array, but the absence of nonlinearity limits overall performance despite good memory capacity (MC = 20.09). This shows that effective reservoir computing requires both memory and nonlinearity. Introducing nonlinearity in the primary cavity alone reduces NMSE to 0.0761, balancing memory and nonlinearity. Adding nonlinearity to the second ring slightly improves NMSE to 0.0635. In Fig.4(b), the second ring's energy is about one-thirtieth of the first, so its contribution is limited. The third ring's energy is one-sixtieth of the second, insufficient to support significant nonlinearity, rendering additional rings effectively linear. The energy in the subsequent micro-rings is even lower, rendering the restriction of their nonlinear behavior inconsequential.

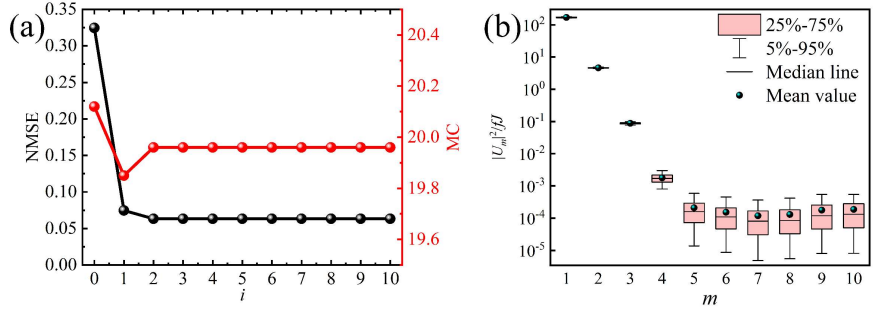

Fig. 4 (a) Performance and memory state of the SCMRRs-9 system when the first $i$ rings (out of 10) exhibit nonlinear behavior. (b) Intracavity energy levels of each ring in the SCMRRs-9 system during training, with no restrictions on nonlinear behavior.

*5.3 symbol duration*

The bit period duration is a key parameter affecting reservoir system performance. In this study, the number of virtual nodes is fixed at 50, ensuring $\tau/\theta=50$, to isolate the impact of bit period on performance. Figures 4(a) and 4(b) show optimal NMSE, MC, and $\sigma(\lambda_0)$ results for various bit periods. The optimal range is 10–60 ns, where system performance remains stable. Below 10 ns, performance degrades as the bit period is too short for effective nonlinear expansion. Above 60 ns, performance also deteriorates as the bit period approaches the thermal relaxation time (100 ns) and the virtual time interval $\theta$ approaches the photon lifetime $\tau_{ph}\approx1845$ps, disrupting short-term memory. Overall, an excessively short bit period reduces nonlinear effectiveness, while a long bit period impairs critical memory. Figures 4(c) and 4(d) show NMSE and MC variations with $Q_1/Q_0$ and MRR count $M$, similar to results at a 10 ns bit period, with MC generally inversely related to NMSE and showing local fluctuations with parameter changes.

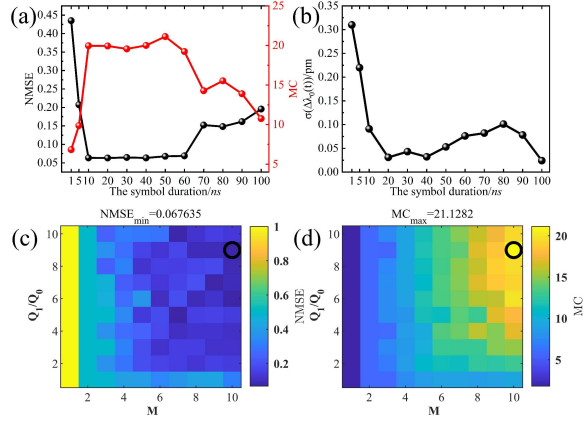

Fig. 5 (a) Relationship between symbol duration, system performance, and memory with 50 nodes. (b) Relationship between symbol duration, and standard deviation of the resonance wavelength shift σ(λ0) with 50 nodes. For a symbol duration of 50 ns, NMSE(c) and MC(d) versus the quality factor's ratio $Q_1/Q_0$ and the total number M of MRRs.

### 5.4 multi-task parallel computing

In this section, the proposed RC system's multi-task computing performance is evaluated using four classic computational tasks: NARMA-10, wireless channel equalization, Mackey-Glass, and Santa Fe. To simplify the simulation process, the system uses the same data length for all four tasks: 1,000 symbols for preheating, 3,000 symbols for training, and 6,000 symbols for testing. The optical input power is optimized between 1 μW and 200 μW, with consistent input power across all task channels. Similarly, the input light detuning is optimized between -3 pm and 3 pm, ensuring consistent detuning across all channels. The resonance wavelength offset between the primary cavity and auxiliary cavity array is also kept consistent, with no nonlinear restrictions imposed on any micro-ring. Finally, due to the inability of light to carry negative signals and the presence of very small negative values in the wireless channel equalization task, the bias $β$ for this task is adjusted from 8 to 30, while it remains unchanged for the other tasks.

In Section 5.2, the lack of nonlinearity in the later micro-rings of the SCMRRs is attributed to insufficient intracavity energy. When multiple input lights resonate in the same primary ring, although at different resonance frequencies, their energy amplitudes are superimposed, suggesting a strategy for reducing input light power. In Fig. 6(a), the optimal optical power for each task is shown as a function of the number of tasks. It is evident that the optimal power decreases as the number of tasks increases, indicating that this approach not only enables simultaneous multi-task computing but also reduces the reservoir system's input power requirements. Figures 6(b)-(e) present the optimized results for each task under the condition of simultaneous computation of four tasks. The thermo-optic and photorefractive effects depend on the modal amplitude contributions of each optical channel. Consequently, each task's input influences the nonlinear dynamics of its respective optical channel while also being affected by the other channels. As shown in Figures 6(b)-(e), all four tasks share a common performance-optimized region (low prediction error or high accuracy) while being constrained to a narrower range of input power. Even without complex optimization across the full 5-D parameter space ($P_M$, $Δλ$s, $Q_1/Q_0$, $M$, offset), the system can still effectively solve these tasks simultaneously with good performance. As indicated in Table 2, each task incurs a slight performance loss relative to its optimal result under these conditions. Nonetheless, the system demonstrates good performance in parallel computing. Future research could further enhance the system's computational performance and versatility by optimizing all parameters and increasing the number of channels.

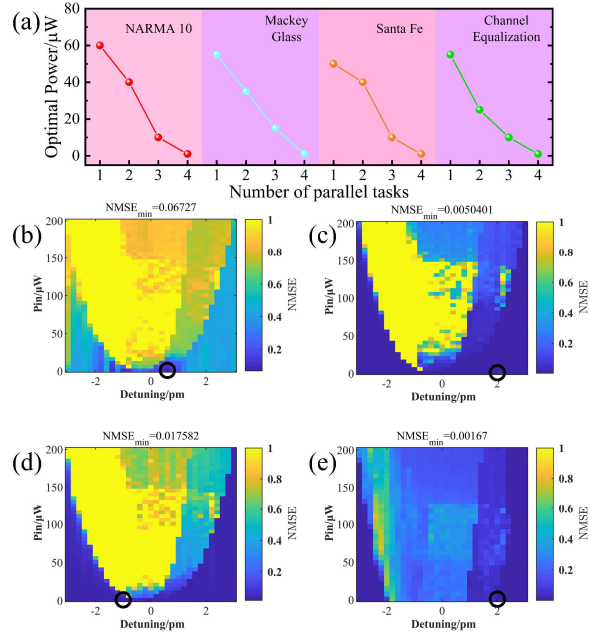

Fig.6 (a) The relationship between the number of different parallel tasks and the power level at which each task achieves the best result. Results of multi-task parallel processing based on WDM for (b) Narma10, (c) Mackey Glass, (d) SantaFe, and (e) Channel Equalization.

**Table 2. The optimal results of each task under different numbers of parallel tasks**

| Number of parallel tasks | Task Type | | | |
|---|---|---|---|---|
| | Narma 10 | Mackey Glass | Santa Fe | Channel Equalization |
| 1 | 0.06345 | -- | -- | -- |
| 1 | -- | 0.0035 | -- | -- |
| 1 | -- | -- | 0.0136 | -- |
| 1 | -- | -- | -- | 0.0010 |
| 2 | 0.0680 | 0.0056 | -- | -- |
| 2 | 0.0682 | -- | 0.0151 | -- |
| 2 | 0.0760 | -- | -- | 0.0010 |
| 2 | -- | 0.0055 | 0.0163 | -- |
| 2 | -- | 0.0052 | -- | 0.0018 |
| 2 | -- | -- | 0.0155 | 0.0015 |
| 3 | 0.0655 | 0.0036 | 0.0141 | -- |
| 3 | 0.0766 | -- | 0.0148 | 0.0010 |
| 3 | 0.0752 | 0.0042 | -- | 0.0012 |
| 3 | -- | 0.0043 | 0.0146 | 0.0013 |
| 4 | 0.06727 | 0.0050 | 0.01758 | 0.0017 |

## 6. Conclusion

In this study, we conducted a numerical investigation of the SCMRR system based on lithium niobate micro-rings. Compared to SCMRR systems based on Si micro-rings, our proposed system demonstrates superior performance, leveraging the excellent nonlinear properties of lithium niobate. We also conducted a detailed analysis of the resonance wavelength offset between the auxiliary cavity array and the primary cavity, the nonlinear behavior of rings within the auxiliary cavity array, the bit period duration, and the system's capability for multi-task simultaneous computation. For the resonance wavelength offset between the auxiliary cavity array and the primary cavity, we identified three distinct regions, each showing very different levels of prediction error. Regarding the nonlinear behavior of the rings in the auxiliary cavity array, we found that the nonlinearity of the first two rings plays a critical role, while the later rings lose their influence on system performance due to insufficient intracavity energy. Concerning the bit period duration, a bit period that is too short hinders the system's timely response, while a period that is too long results in short-term memory loss. Thus, there exists an optimal range for the bit period under a fixed number of nodes. For multi-task simultaneous computation, introducing multiple WDM channels to the original SCMRR system creates a multi-task time-delay RC system capable of solving four traditional TDRC tasks simultaneously with good performance. While we limited the number of parallel tasks to four, using more WDM channels could enable a higher degree of parallelization. Additionally, the slower relaxation speeds of lithium niobate's nonlinear effects, compared to Si, result in reduced computational speed; however, the multi-task parallel approach effectively mitigates this issue. Leveraging existing manufacturing technologies, the proposed RC system paves the way for scalable, integrated photonic RC systems.

**Funding.** National Natural Science Foundation of China (60907032, U23A2074); Natural Science Foundation of Zhejiang Province (LZ24F050008, LY20F050009); Open Fund of the State Key Laboratory of Advanced Optical Communication Systems and Networks (2020GZKF013); Horizontal projects of public institution (KY-H-20221007, KYY-HX-20210893).

**Acknowledgments.** This work was partially carried out at the USTC Center for Micro and Nanoscale Research and Fabrication.

**Disclosures.** The authors declare that there are no conflicts of interest related to this article.

**Data availability.** Data underlying the results presented in this paper are not publicly available at this time but may be obtained from the authors upon reasonable request.